\documentclass{elsart}
\usepackage{epsfig}

\begin{document}

\begin{frontmatter}

\title{Directed cycles and related structures in random graphs: I- Static properties}

\author[vcb]{Valmir C. Barbosa}
\ead{valmir@cos.ufrj.br}
\author[rdsrs]{Raul Donangelo}
\ead{donangel@if.ufrj.br}
\author[rdsrs]{Sergio R. Souza}
\ead{srsouza@if.ufrj.br}

\address[vcb]
{Programa de Engenharia de Sistemas e Computa\c c\~ao, COPPE,\\ 
Universidade Federal do Rio de Janeiro,\\
C.P. 68511, 21941-972 Rio de Janeiro, Brazil}
\address[rdsrs]
{Instituto de F\'\i sica, Universidade Federal do Rio de Janeiro,\\ 
C.P. 68528, 21941-972 Rio de Janeiro, Brazil}

\begin{abstract}
We study directed random graphs (random graphs whose edges are directed), and
present new results on the so-called strong components of those graphs. We
provide analytic and simulation results on two special classes of strong
component, called cycle components and knots, which are important in random
networks that represent certain computational systems.
\end{abstract}


\end{frontmatter}

\section{Introduction}\label{intr}

Networks formed by units which interact with one another are found in all
sorts of fields, ranging from sociology and economics to biology and
physics. Some of these systems exhibit spatial order, such as crystal
lattices in solid state physics. Most do not have such order, and 
constitute the so-called random networks. 
Probably the most well known example of such systems is the World Wide Web
(WWW). Its randomness arises from the fact that any user
may create a web page with an arbitrary number of connections to other
web pages. Many similar examples exist in ecology, industry,
and transportation, to name just a few areas in which such systems have
received considerable attention in recent years.

It is surprising, considering the importance of these networks in
our daily lives, that so much remains unknown about their properties,
especially their topological structures. One notable exception is
the case of scientific collaborations, which will certainly be of interest to
the readers of this work. These collaborations have been extensively studied by
Newman \cite{n01a,n01b,n01c}, among others, but this has only been possible
because such collaborations, like those of
co-actors in movies or co-directors in prominent companies, are well 
documented. It has been found that the average number of connections per
node increases linearly with the number of nodes, thus the networks
become denser and, in particular, the average shortest path
between nodes decreases with time.

The difficulty in studying several of the other real
networks lies in the unavailability
of data on them. This is in part due to the fact that these 
are very large, evolving networks---for some of them, the evolution runs at a 
very fast pace, as is the case of the WWW. But the situation is changing 
very rapidly due to the new methods of data acquisition applied to 
the various networks existing in real life. A major step in the
understanding of the structure of large networks was recently taken by
Barab\'asi and collaborators \cite{ba99,baj00,ab00},
who unveiled the large degree
of self-organization in complex networks. They explored several large 
databases containing information on networks spanning such diverse 
fields as the WWW or scientific citations, and found that for all 
these systems the number of connections between nodes in the network 
follows a power-law behavior, with exponents ranging from about $-2$ to 
$-4$, depending on the particular system considered. We refer the reader to the
recent review by these authors \cite{ab02} for details.

The structure of random static networks was discussed over four
decades ago, in the seminal work of Erd\H{o}s and R\'{e}nyi \cite{er59}.
One should also mention that these same authors started the study of
evolving networks \cite{er60}. 
The static structure of networks has been studied employing graph
theory \cite{jlr00}. Most results are for the case of a Poisson
distribution of connections; recently, however, results have also been
obtained for networks with arbitrary distributions \cite{mr95,mr98}.

Some of these graphs have directed connections, such as the links
between web pages or scientific journal citations; such graphs are referred
to as digraphs. Others are not directed, e.g., collaborations in science or
other activities. In this work we concentrate on the most general case of
directed networks, starting with the case of static connections. The study
of evolving directed networks will be the subject of a separate
publication \cite{bds02}. We mention, incidentally, that an evolving undirected
network that self-organizes into a critically connected one has already been
presented by one of us and collaborators in \cite{cdks98}.

This paper is organized as follows. In Section \ref{old} we discuss some
properties of random digraphs, obtained both analytically and through
computer simulations. Different types of topological structures
found in those graphs and their properties are studied in Section \ref{new}.
In the last section we draw conclusions from the results obtained in
the present work.

\section{Random digraphs}\label{old}

\subsection{Analytic results}\label{old:analytic}

We first discuss random (undirected) graphs, since several of their
properties carry over almost directly to the case of random directed graphs, or
random digraphs.

We consider random graphs on the fixed set of nodes $N=\{1,\ldots,n\}$
within the so-called constant edge-probability model \cite{b01}. If
$G$ is such a graph, then an edge exists between nodes $i$ and $j$ (including
$i=j$) in $G$ with probability $p$, independently of $i$ or $j$. Multiple
edges joining the same pair of nodes are not allowed, so $1-p$ is the
probability that $i$ and $j$ are not connected by an edge. When two nodes
are joined by an edge, we call them neighbors in the graph. If we let the
expected number of neighbors of a node (the node's degree) be denoted by $z$,
then clearly $z=pn$.

Because edges are present independently of all other node pairs, the
probability that a node has degree $k$ is given by
\begin{equation}
\label{degreeprob}
{n\choose k}p^k(1-p)^{n-k},
\end{equation}
for $k\ge 0$. For large $n$, (\ref{degreeprob}) can be approximated as in
\begin{equation}
\label{poisson}
\frac{z^k}{k!}\left(1-\frac{z}{n}\right)^n\approx\frac{z^ke^{-z}}{k!},
\end{equation}
which is the Poisson distribution with mean $z$, denoted by $P(z,k)$.

Several properties are known to hold for $G$ when $n$ is large \cite{b01}.
Of special interest are the properties that describe the connected components of
$G$. A subgraph $G'$ of $G$ is a connected component if its node set, call it
$N(G')$, is maximal with respect to the property that any two of its nodes can
be reached from each other by traversing edges. Maximal here means that this
reachability property does not hold for any other node set that strictly
contains $N(G')$.

When $z<1$, the expected number of nodes in the connected component to which a
randomly chosen node, say $i$, belongs can be assessed as follows. First we
count $i$ itself, then its expected number of neighbors, then its neighbors'
expected number of neighbors, and so on; for $n$ large, the probability that any
two of these nodes are the same node is nearly zero, so the number we seek tends
to
\begin{equation}
\label{zlesst1}
\sum_{k\ge 0}z^k=\frac{1}{1-z}.
\end{equation}

\begin{figure}
\centering
\epsfig{file=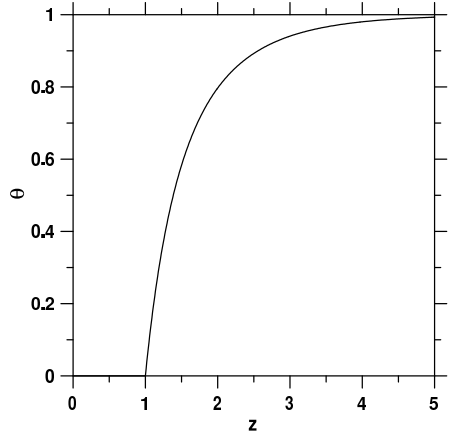, height=2.25in}
\caption{$\theta$ as a function of $z$.}
\vspace{0.30in}
\label{thetaXz}
\end{figure}

By (\ref{zlesst1}), the expected size of a connected component remains finite
even when $n$ tends to infinity, so long as $z<1$. One natural question is
whether finite-size connected components can still be expected to exist for
infinite $n$ when $z>1$. In order to answer this question, we resort to the
Poisson approximation of (\ref{poisson}) and describe the process of discovering
the nodes that are reachable from $i$ as the following Poisson branching
process: node $i$, the progenitor, spawns a Poisson-distributed
number of nodes, and similarly for each newly spawned node. If $q$ is the
probability that a node's total progeny in this process is finite, then it
must satisfy the condition
\begin{equation}
q=\sum_{k\ge 0}\frac{z^ke^{-z}}{k!}q^k,
\end{equation}
yielding the transcendental equation
\begin{equation}
q=e^{-z(1-q)},
\end{equation}
or, for $\theta=1-q$,
\begin{equation}
\label{transcend}
1-\theta=e^{-z\theta}.
\end{equation}
The unique root of (\ref{transcend}) in $[0,1]$ depends on $z$ as shown in
Figure \ref{thetaXz}.

For $z>1$ and $n$ approaching infinity, the size of the connected component to
which $i$ belongs is then as follows: with probability $\theta$ it tends to
infinity (also said to be large); with probability $1-\theta$ it is finite
(also said to be small). If it is small, then the expected number of neighbors
of node $i$ is $z(1-\theta)$ and, similarly to the reasoning that led to
(\ref{poisson}), it can be shown that the probability that node $i$ has degree
$k$ is $P(z(1-\theta),k)$. Not only this, but since $z(1-\theta)$ can be easily
shown to be less than $1$, we can retrace the steps that led to (\ref{zlesst1})
and conclude that the expected size of a small component is
\begin{equation}
\label{zgreatert1}
\sum_{k\ge 0}[z(1-\theta)]^k=\frac{1}{1-z(1-\theta)}.
\end{equation}
In addition, it can also be shown that every small connected component of $G$
(for both the $z<1$ and the $z>1$ cases) comprises no more than $A\ln n$ nodes,
where $A$ is a constant \cite{k90}.

If the component is large, then it is necessarily the only large component in
the graph, called the giant component, and its size is $\theta n$. So, in the
vicinity of $z=1$, a sharp phase transition takes place and is characterized
by the appearance of a giant connected component. The appearance of this
component has been studied in detail through the use of a generating-function
formalism \cite{jklp93}, but the simple characterization based on a
Poisson branching process we have seen in this section is from
\cite{k90}.

We now discuss some of the known properties of random digraphs. If $D$ is a
random digraph, then in the constant edge-probability model an edge exists from
node $i$ to node $j$ in $D$ with probability $p$, independently of $i$ or $j$
and including the case $i=j$. Note that, unlike the undirected case, it is now
possible for two edges to exist between $i$ and $j$, since the edge from $j$ to
$i$ occurs independently with probability $p$ as well. Except for this one case,
edge multiplicity between the same two nodes is disallowed. In a random digraph,
a node $i$ has two degrees, the in-degree (number of nodes $j$ such that an edge
from $j$ to $i$ exists) and the out-degree (number of nodes $j$ such that an
edge from $i$ to $j$ exists). As in the undirected case, both degrees have the
same expected value, denoted by $z=pn$, and for large $n$ can be assumed to be
distributed as $P(z,k)$.

In the case of digraphs, the reachability properties among nodes depend on the
directions of the edges, so the notion of a connected component gives way to
more specialized structures. Two of these structures are the in-component and
the out-component to which node $i$ belongs; these are, respectively, the
subgraph of $D$ from whose nodes $i$ can be reached by traversing edges along
their directions, and the subgraph of $D$ whose nodes can be reached from $i$
likewise. For large $n$, both in-components and out-components are amenable to
the same approximations used to analyze the connected components of random
graphs, which results in equivalent properties \cite{k90}. Small in- and
out-components are the rule for $z<1$ but also occur for $z>1$ with probability
$1-\theta$; as the case may be, the expected size of such a component is given
by either (\ref{zlesst1}) or (\ref{zgreatert1}), but the actual size is, with
very high probability, bounded by $A\ln n$. For $z>1$, with probability $\theta$
node $i$ belongs to the giant in-component of size $\theta n$, the same holding
for the giant out-component.

But the most important structure arising in $D$ as a consequence of edge
directionality is the strongly connected component, or simply strong component.
A subgraph $D'$ of $D$ is a strong component if its node set, call it $N(D')$,
is maximal with respect to the property that, for every two nodes $i$ and $j$
in $N(D')$, both $j$ can be reached from $i$ and $i$ from $j$, always
respecting the edges' directions.

\begin{figure}
\centering
\begin{tabular}{c@{\hspace{0.65in}}c}
\epsfig{file=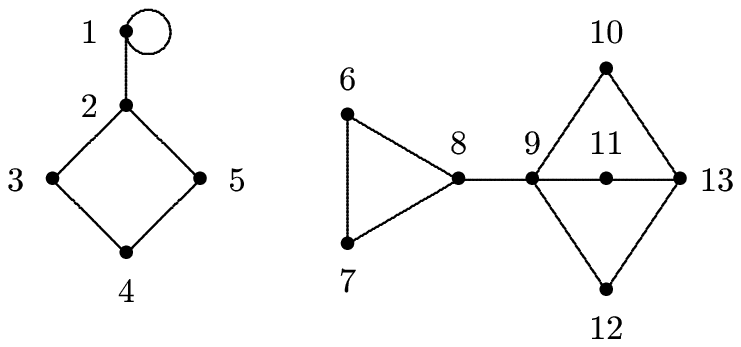, width=2.33in}&\epsfig{file=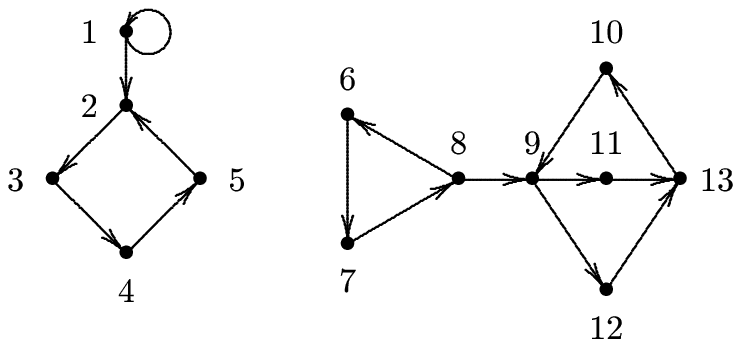, width=2.33in}\\
\vspace{-0.1in}
{\footnotesize (a)}&{\footnotesize (b)}\\\\
\end{tabular}
\begin{tabular}{c}
\epsfig{file=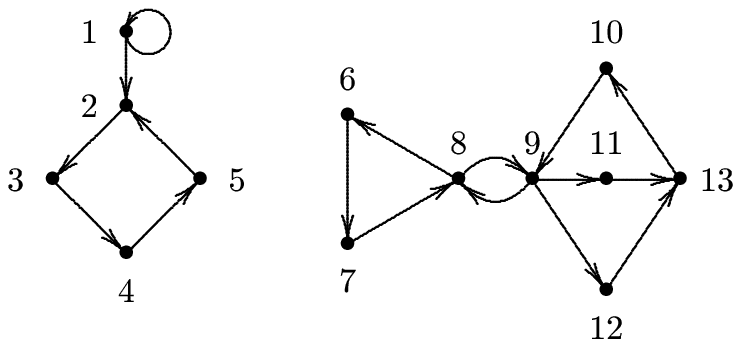, width=2.33in}\\
\vspace{-0.1in}
{\footnotesize (c)}\\\\
\end{tabular}
\caption{A graph with two connected components (a), a digraph with four strong
components (b), and a digraph with three strong components (c).}
\vspace{0.30in}
\label{components}
\end{figure}

We give in Figure \ref{components} an illustration
of the notion of a strong component. Figure \ref{components}(a) shows a
$13$-node graph with two connected components, while Figure \ref{components}(b)
shows a digraph obtained from the graph of Figure \ref{components}(a) by the
assignment of directions to the edges. This digraph has four strong components,
with node sets $\{1\}$, $\{2, 3, 4, 5\}$, $\{6, 7, 8\}$, and
$\{9, 10, 11, 12, 13\}$. Figure \ref{components}(c) depicts yet another digraph
on the same node set, this one comprising all the edges of
Figure \ref{components}(b) and the additional edge from node $9$ to node $8$.
Notice that the addition of this single edge causes two of the strong
components of Figure \ref{components}(b) to be merged into the single strong
component of node set $\{6, 7, 8, 9, 10, 11, 12, 13\}$.

For the random digraph $D$, results similar to the ones we have seen for
in- and out-components hold for the strong components \cite{k90}. If
$z<1$, then strong components are all small and have sizes bounded by
$A\ln n$. If $z>1$, the same holds for all strong components but one, the
giant strong component, whose size is $\theta^2n$. The latter result is
essentially a consequence of the following properties. The events
``the in-component of node $i$ is large'' and ``the out-component of node $i$
is large'' are nearly independent, and from this it can be shown that a set $S$
of nodes whose members all have large in- and out-components has size
$\theta^2n$. Also, node $j$ is reachable from node $i$ if both the in-component
of $j$ and the out-component of $i$ are large, so $S$ is a strong component, as
every node not in $S$ belongs to a small strong component.

\subsection{Results of simulations}\label{old:simulation}

\begin{figure}
\centering
\begin{tabular}{c@{\hspace{0.3in}}c}
\epsfig{file=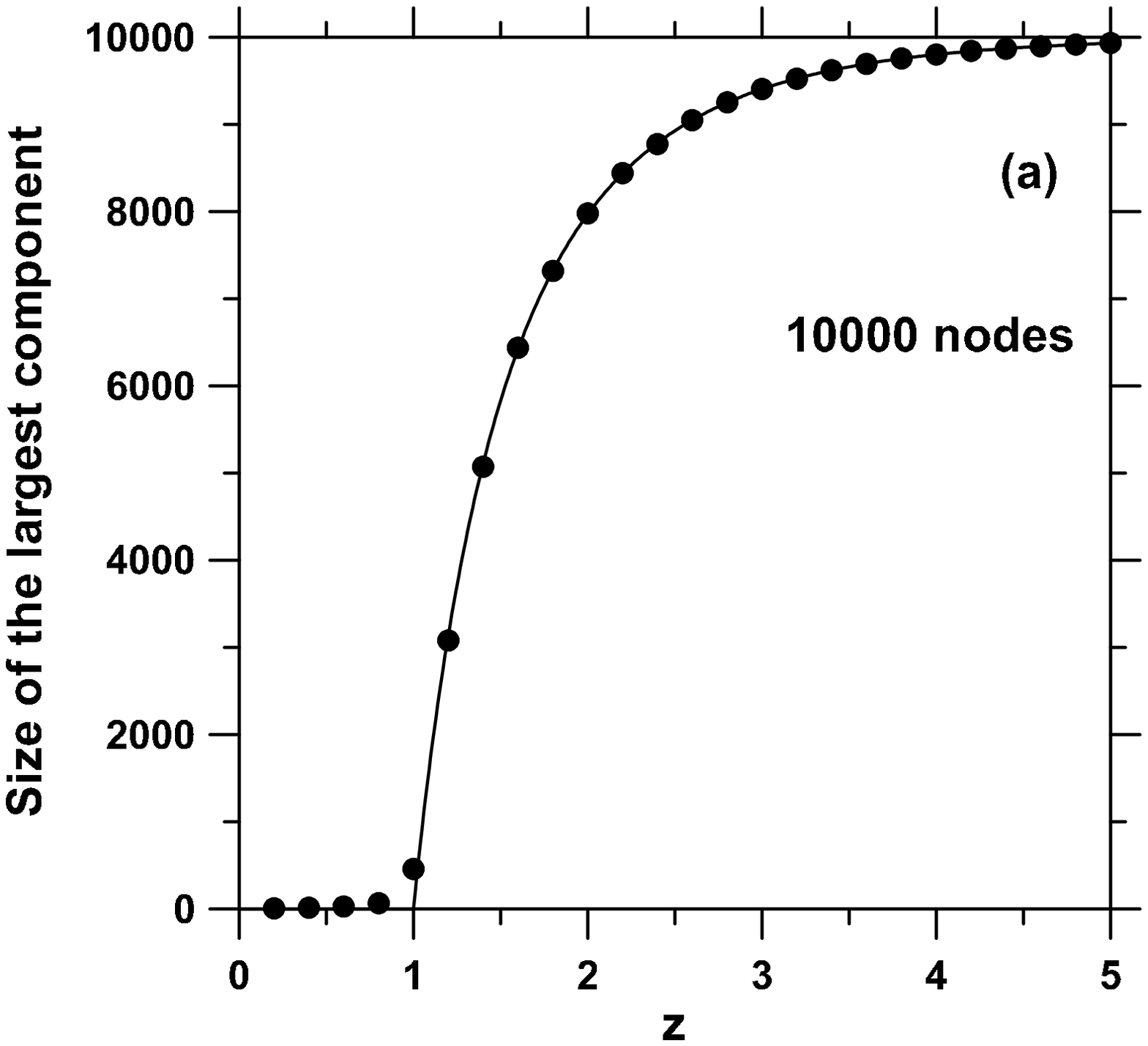, height=2.25in}&\epsfig{file=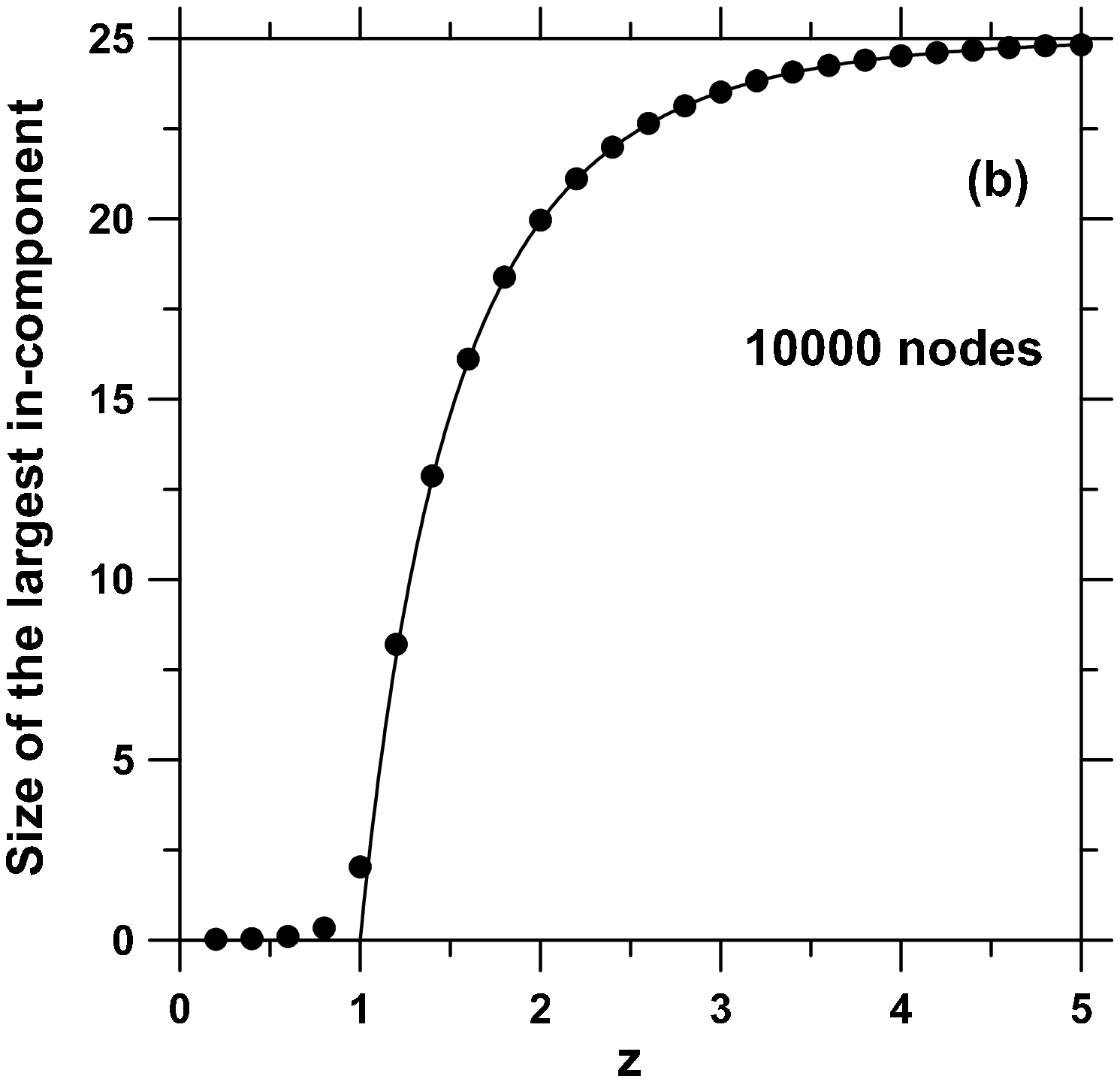, height=2.25in}\\
\epsfig{file=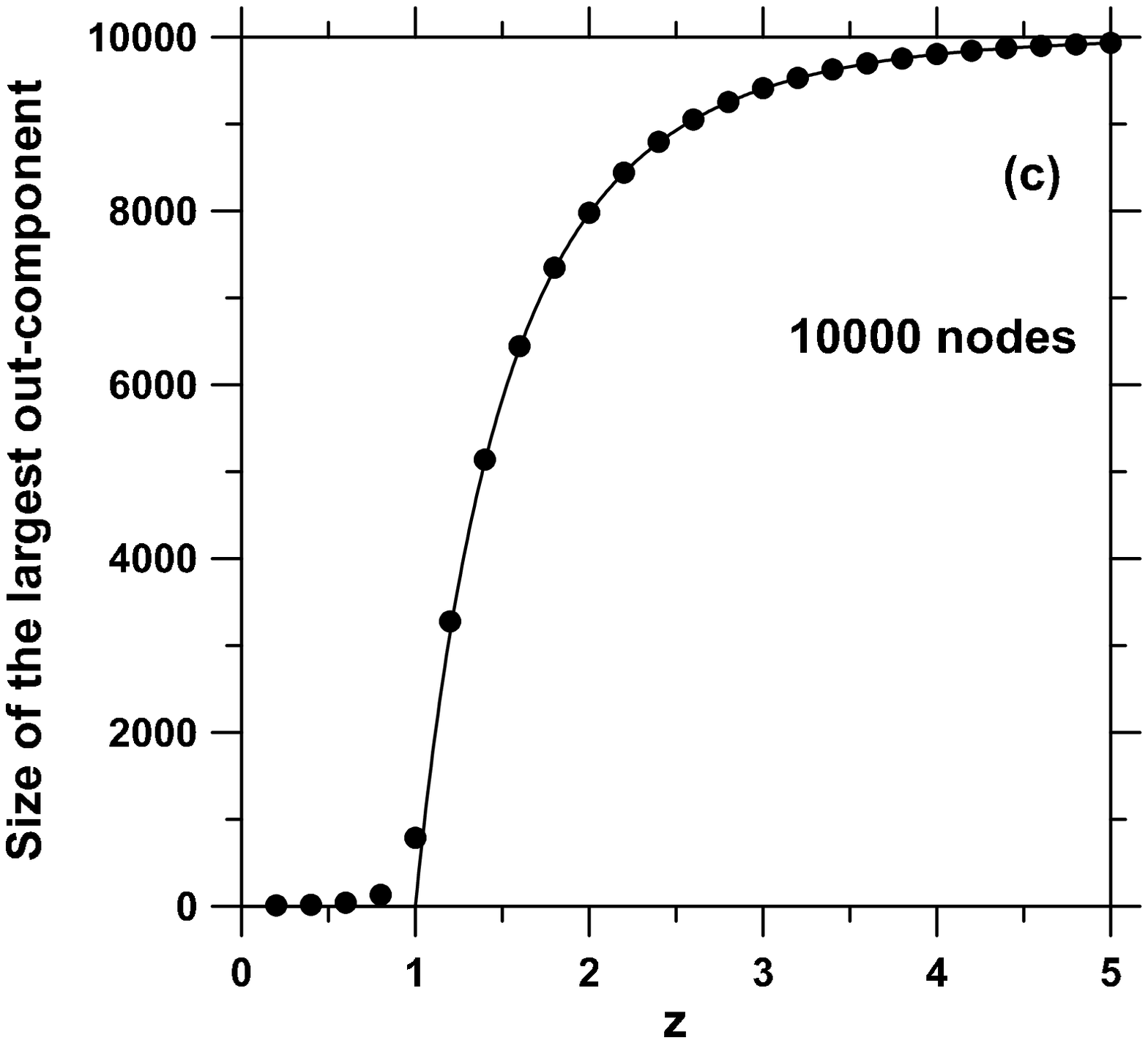, height=2.25in}&\epsfig{file=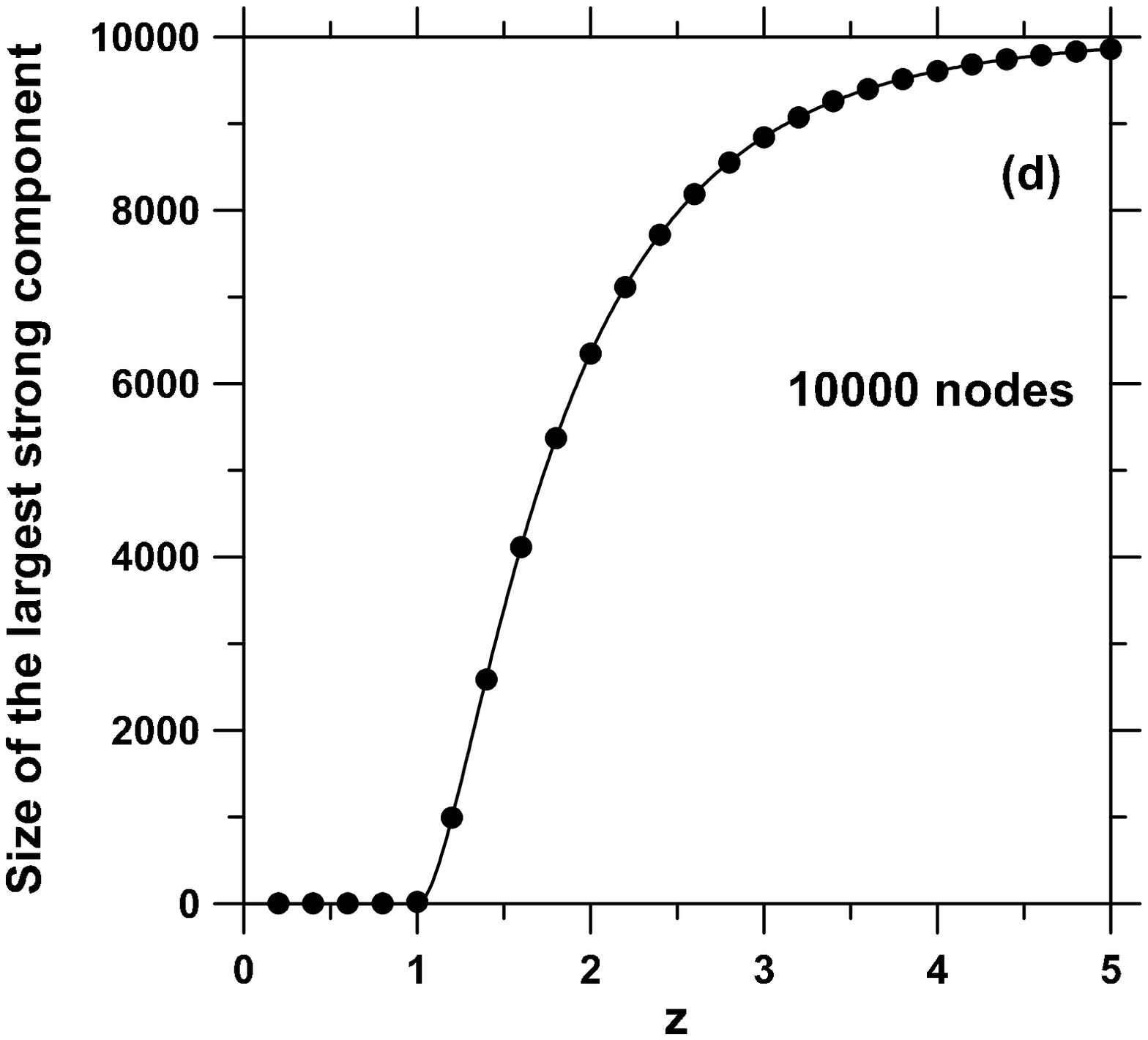, height=2.25in}\\
\end{tabular}
\caption{Average size of the largest connected component (a), the largest
in-component (b), the largest out-component (c), and the largest strong
component (d) as a function of $z$ ($100$ simulation runs). For $z>1$, each of
these sizes refers to the unique giant component of the corresponding type.
Continuous lines are plots of the analytic predictions, also for $z>1$.
Simulation results are indicated by filled circles.}
\vspace{0.30in}
\label{giantXz}
\end{figure}

In this section we illustrate the main analytic results discussed in Section
\ref{old:analytic} by presenting the results of computer simulations.
For $n$ and $z$ fixed, each simulation run first generates a random graph
(or digraph) and then applies to it a procedure based on depth-first search
\cite{clrs01} to find all of its components (or strong components). This is
repeated several times and the quantities of interest are averaged over the
repetitions. Although each run of the simulation performs relatively
efficiently (it requires $O(zn)$ time for completion), for large $n$ the number
of runs is limited by the available computational resources. The results we
present in this section were obtained for $n=10000$ over $100$ independent
runs. They are given in Figure \ref{giantXz} for $z\le 5$ along with plots of
the analytic predictions.

The appearance of the giant component in a random graph as $z$ is increased
through the $z=1$ boundary is shown in Figure \ref{giantXz}(a). For random
digraphs, Figures \ref{giantXz}(b) and \ref{giantXz}(c) depict the appearance of
the giant in-component and the giant out-component, respectively. As expected,
the three phenomena are entirely equivalent to one another, and for $z>1$ comply
with the $\theta n$ prediction for the components' expected sizes.

Figure \ref{giantXz}(d) also corresponds to random digraphs and shows the
appearance of the giant strong component as $z$ is increased through $z=1$.
A comparison to Figures \ref{giantXz}(a) through \ref{giantXz}(c) reveals
that the plots are now slightly shifted to the right, which is in accordance
with the prediction of $\theta^2n$ when $z>1$ for the size of the giant strong
component.

\section{Structures in random digraphs}\label{new}

\subsection{Analytic results}\label{new:analytic}

Our primary concern in this paper is to further the study of the strong
components of random digraphs. Our specific interest is the study of the
so-called cycle components and knots. A cycle component is a strong component
whose nodes are arranged in a single directed cycle, as for example the
components of node sets $\{1\}$, $\{2, 3, 4, 5\}$, and $\{6, 7, 8\}$ in
Figure \ref{components}(b). A knot is a strong component whose nodes'
out-components all coincide with the strong component itself; from a node
inside a knot, it is impossible to reach any node outside the knot by following
edges along their directions. Figure \ref{components}(b) contains examples of
knots as well, specifically the strong components of node sets $\{2, 3, 4, 5\}$
and $\{9, 10, 11, 12, 13\}$. Similarly for the strong component of node set
$\{6, 7, 8, 9, 10, 11, 12, 13\}$ in Figure \ref{components}(c).

Cycle components and knots are of interest because they are among the digraph
structures of greatest interest in the context of the so-called safety
conditions of distributed computing \cite{l96}. Some of these conditions
relate to the prohibition of deadlocks, that is, situations involving a group of
computational processes in which all processes are blocked waiting for an
action to be taken by some process of the very group. Such situations are
clearly undesirable, since they lead part of the computational system (or even
all of it) to a global state of wait for a condition that can never be satisfied
and thus calls for outside intervention. Digraphs can be used to model the
computational system and the various waits involved, and cycle components and
knots are some of the structures to be dealt with \cite{b02}.

We discuss cycle components first. In our study, we shall need the distribution
of the collective total progeny of a group of $m$ nodes when each of them
generates its individual progeny by the Poisson branching process discussed in
Section \ref{old:analytic}. The case of $m=1$ asks for the total-progeny
distribution whose expected value is given as in (\ref{zlesst1}) for $z<1$ and
as in (\ref{zgreatert1}) for $z>1$. In general, for $\lambda<1$ the total
progeny of a $P(\lambda,k)$ branching process is known to be distributed as
the Borel distribution with mean $1/(1-\lambda)$ \cite{a98}. According to this
distribution, the probability that a node's total progeny has size $k$ is
$B(\lambda,k)$, given for $k\ge 1$ by
\begin{equation}
\label{borel}
B(\lambda,k)=\frac{(\lambda k)^{k-1}e^{-\lambda k}}{k!}.
\end{equation}

One possible approach to the derivation of (\ref{borel}) is the one described
in \cite{gk91}. It is particularly instructive in our context, because it
can be applied to the case of $m>1$ as well; we discuss it briefly before
proceeding. The idea is to start with the generating function for the Borel
distribution,
\begin{equation}
\label{genborel1}
\mathcal{B}(x)=\sum_{k\ge 1}B(\lambda,k)x^k.
\end{equation}
Next, by well-known properties of probability generating functions
\cite{f68,gkp94}, and considering that the generating function for
the Poisson distribution is
\begin{equation}
\sum_{k\ge 0}P(\lambda,k)x^k=e^{\lambda(x-1)},
\end{equation}
we see that $\mathcal{B}(x)$ has to satisfy the relation
\begin{equation}
\label{genborel2}
\mathcal{B}(x)=xe^{\lambda(\mathcal{B}(x)-1)},
\end{equation}
where the $x$ factor accounts for the need to compensate for the summation from
$k=1$ in (\ref{genborel1}), instead of $k=0$, and the exponential gives the
generating function for the distribution of the sum of a Poisson-distributed
number of independent, Borel-distributed random variables.

Solving for $\mathcal{B}(x)$ directly from (\ref{genborel2}) for later
computation of $B(\lambda,k)$ by differentiation is usually not feasible, but
if we use $b$ for $\mathcal{B}(x)$, rewrite (\ref{genborel2}) as
\begin{equation}
x=f(b)=be^{-\lambda(b-1)},
\end{equation}
and let
\begin{equation}
g(b)=b,
\end{equation}
we see that Lagrange's expansion \cite{as65} of $g$ can be applied
directly: for $f'(0)\neq 0$, Lagrange's expansion gives an expression
for $g$ as the power series in $x$,
\begin{equation}
\label{lagrange}
g(b)=g(0)+\sum_{k\ge 1}\frac{x^k}{k!}
\left[
\frac{d^{k-1}}{db^{k-1}}\left(g'(b)\left(\frac{b}{f(b)}\right)^k\right)
\right]_{b=0},
\end{equation}
so long as $g(b)$ is infinitely differentiable. In order for the expression in
(\ref{lagrange}) to equal the one in (\ref{genborel1}), it suffices for the
coefficient of $x^k$ in (\ref{lagrange}) to be equal to $B(\lambda,k)$ as
given in (\ref{borel}). This can be easily verified.

Let us then generalize beyond the $m=1$ case. For $m\ge 1$ and $k\ge m$, let
$S(m,\lambda,k)$ be the probability that the Borel-distributed progenies of $m$
nodes add up to $k$. If we assume independence of the Poisson branching
processes that generate those progenies (this is certain to hold as $n$ tends
to infinity, as we argued in Section \ref{old:analytic}), then $S(m,\lambda,k)$
is generated by
\begin{equation}
\mathcal{B}(x)^m=\sum_{k\ge m}S(m,\lambda,k)x^k,
\end{equation}
once again by well-known properties of probability generating functions.

If we now let
\begin{equation}
g(b)=b^m,
\end{equation}
then we find that the coefficient of $x^k$ in (\ref{lagrange}) is
\begin{equation}
\label{sumborel}
S(m,\lambda,k)=\frac{m/k}{(k-m)!}(\lambda k)^{k-m}e^{-\lambda k}
\end{equation}
for $k\ge m$ (for $1\le k<m$, the coefficient is found to be zero). Naturally,
$S(1,\lambda,k)=B(\lambda,k)$.

We now turn to the cycle components of a random digraph $D$. Our aim is to
compute the expected number of cycle components for a given value of $z$. We
start with the case of $z<1$, letting $\kappa_{z<1}(z)$ denote the expected
number of cycle components in this case. When $z<1$, we know that no strong
component has more than $A\ln n$ nodes, where $A$ is a constant, so this is also
an upper bound on the number of nodes in a cycle component.

Given a fixed set $M$ of $m$ nodes and a fixed circular arrangement $\alpha$ of
those nodes, the probability $\pi(z,m)$ that a cycle component exists having
$M$ for node set and edges in conformity with that circular arrangement depends
on the nodes' in- and out-components. The only possibility for large $n$ is that
all such components are small and can be modeled by the mutually independent
Poisson branching processes we have been considering. So, conditioning on this
possibility, we see that the probability we seek is simply the probability $p^m$
that $m$ directed edges are present connecting consecutive nodes of $M$ in the
order established by $\alpha$. We uncondition by multiplying $p^m$ by
the probability of small, collective total progenies both in the process that
generates the in-components and the out-components. Because the $m$
in-components' combined total progeny is at most $A\ln n$, and using
(\ref{sumborel}) with $\lambda=z$ for the probability of the combined progenies
of $m$ independent Poisson branching processes, we get, with $z/n$ for $p$,
\begin{equation}
\label{prob}
\pi(z,m)=\left(\frac{z}{n}\right)^m
\sum_{k=m}^{A\ln n}S(m,z,k)\sum_{\ell=m}^{A\ln n}S(m,z,\ell).
\end{equation}

The final expression for $\kappa_{z<1}(z)$ is obtained by considering that there
are \raise.5ex\hbox{$\scriptscriptstyle{n\choose m}$} possibilities for the set
$M$ and that, of the $m!$ possibilities for $\alpha$ given $M$, only one out of
$m$ counts (the others represent different starting nodes on the same circular
arrangement). We then get for the expected number of cycle components when
$z<1$,
\begin{equation}
\label{nczlesst1}
\kappa_{z<1}(z)=\sum_{m=1}^{A\ln n}
{n\choose m}(m-1)!\;\pi(z,m).
\end{equation}

For $z>1$, let $\kappa_{z>1}(z)$ denote the expected number of cycle components.
The first fact to note in this case is that the probability that the giant
strong component is a cycle component tends to zero as $n$ tends to infinity.
To see this, momentarily regard all edges as undirected. This might in principle
cause edge duplicity between some pairs of nodes, but for large $n$ the
probability that this happens is clearly negligible. So in the resulting
undirected graph, call it $G(D)$, the probability that two given nodes are
joined by an edge is still $z/n$. The giant strong component is necessarily
embedded in the giant component of $G(D)$, so in $G(D)$ no edges exist joining
any of the $\theta^2n$ nodes of the giant strong component to any of the
$n-\theta n$ edges outside the giant component of $G(D)$. Consider, in addition,
that those $\theta^2n$ nodes are connected to one another by exactly $\theta^2n$
edges. The probability of such a scenario, which is necessary for the giant
strong component to be a cycle component, is
\begin{equation}
\left(\frac{z}{n}\right)^{\theta^2n}
\left(1-\frac{z}{n}\right)^{(\theta^4-\theta^3+\theta^2)n^2-\theta^2n},
\end{equation}
and clearly vanishes for large $n$.

The possible cases for a cycle component in the $z>1$ case are then only three:
(i) all nodes in the cycle component have small in-components and small
out-components; (ii) they all have small in-components but large out-components;
(iii) they all have large in-components but small out-components.

We analyze each of cases (i) through (iii) separately for $M$ and $\alpha$ given
as earlier. Case (i) is entirely analogous to the case of $z<1$, the only
differences being that $\lambda=z(1-\theta)$ in the Poisson branching processes
and that the unconditioning probability that yielded $\pi(z,m)$ must include a
factor accounting for the probability that all the $m$ nodes have small in- and
out-components. This probability is $(1-\theta)^{2m}$, so letting $\pi_1(z,m)$
be the probability of interest in this case we get
\begin{eqnarray}
\label{prob1}
\lefteqn{\pi_1(z,m)=}\nonumber\\
&&\quad\left(\frac{z(1-\theta)^2}{n}\right)^m
\sum_{k=m}^{A\ln n}S(m,z(1-\theta),k)
\sum_{\ell=m}^{A\ln n}S(m,z(1-\theta),\ell).
\end{eqnarray}

Cases (ii) and (iii) are entirely analogous to each other, so we discuss case
(ii) only. This case, too, has similarities to the case of $z<1$: as for case
(i), we need to set $\lambda=z(1-\theta)$ and to fix the unconditioning
probability. But now this probability no longer involves the double summation
of (\ref{prob}), but one single summation instead, corresponding to the small
in-components. The remainder of the necessary fix goes as follows. All $m$ nodes
have small in-components and large out-components. The probability that they all
have small in-components is $(1-\theta)^m$. In order to assess the probability
that they all have large out-components, consider that the only case in which
this does not happen is the case in which they all have small out-components
(in other words, if one of the nodes has a large out-component, then they all
do). The probability that we need is then $1-(1-\theta)^m$. So, for cases (ii)
and (iii) combined, we have, letting $\pi_2(z,m)$ be the probability of
interest,
\begin{equation}
\label{prob2}
\pi_2(z,m)=2\left(\frac{z(1-\theta)}{n}\right)^m[1-(1-\theta)^m]
\sum_{k=m}^{A\ln n}S(m,z(1-\theta),k).
\end{equation}

From (\ref{prob1}) and (\ref{prob2}), and once again considering the different
possibilities for $M$ and for $\alpha$, we obtain
\begin{equation}
\label{nczgreatert1}
\kappa_{z>1}(z)=\sum_{m=1}^{A\ln n}
{n\choose m}(m-1)!\;[\pi_1(z,m)+\pi_2(z,m)]
\end{equation}
for the expected number of cycle components when $z>1$.

We now turn to a discussion of knots in $D$, starting with the probability
that a small strong component of size $m$ is a knot. We know from Section
\ref{old:analytic} that every small component, for $z<1$ and $z>1$ alike, has
at most $A\ln n$ nodes, so this probability is
\begin{equation}
\left[\left(1-\frac{z}{n}\right)^{n-m}\right]^m
\le\left(1-\frac{z}{n}\right)^{An\ln n-(A\ln n)^2},
\end{equation}
which clearly tends to zero for large $n$.

The probability that a large strong component is a knot can be assessed
likewise, but this of course only makes sense for $z>1$, the large strong
component being the unique giant strong component. In this case, the number
of nodes is $\theta^2n$ and the probability, which we denote by $\pi_n(z)$, is
\begin{equation}
\label{bigknots}
\pi_n(z)=\left[\left(1-\frac{z}{n}\right)^{n-\theta^2n}\right]^{\theta^2n}
=\left(1-\frac{z}{n}\right)^{\theta^2n^2(1-\theta^2)}.
\end{equation}

\begin{figure}
\centering
\epsfig{file=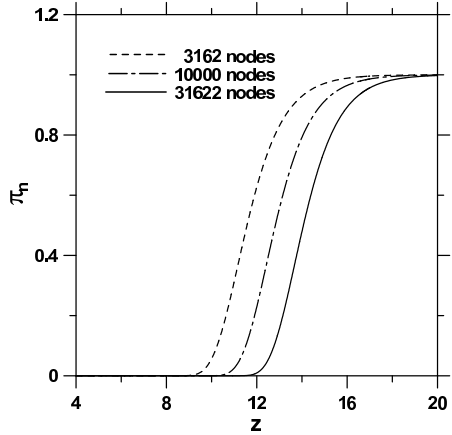, height=2.25in}
\caption{$\pi_n$ as a function of $z$.}
\vspace{0.30in}
\label{piXz}
\end{figure}

We give in Figure \ref{piXz} plots of $\pi_n(z)$ for $n=3162$
($\approx 10^{3.5}$), $n=10^4$, and $n=31622$ ($\approx 10^{4.5}$). What the
curves reveal is that $\pi_n(z)$ tends to zero for small $z$, tends to one for
large $z$, and between the two extremes undergoes a sharp transition from zero
to one at a value of $z$ that seems to increase logarithmically with $n$. In
fact, all these properties follow from (\ref{bigknots}), as we discuss next.

For $z\ll n$, (\ref{bigknots}) can be rewritten as
\begin{equation}
\label{bigknots1}
\pi_n(z)\approx e^{-zn\theta^2(1+\theta)(1-\theta)}.
\end{equation}
Let us then consider three distinct possibilities for $z$, beginning with two
extreme ones. First we let $z$ be of the order of a few units; in this case,
$\theta<1$ and the exponent of (\ref{bigknots1}) tends to infinity along with
$n$, thus leading $\pi_n(z)$ to approach zero. The second extreme possibility
is that of a very large value for $z$, in which case $\theta\approx 1$.
Recalling (\ref{transcend}) and substituting $\ln(1-\theta)$ for $-z\theta$ in
the exponent of (\ref{bigknots1}) yields
\begin{equation}
\pi_n(z)\approx e^{n\theta(1+\theta)(1-\theta)\ln(1-\theta)},
\end{equation}
where the exponent is seen to tend to zero and $\pi_n(z)$ to one.

The third possibility we consider for $z$ is the value $\bar z$ for which
$\pi_n(\bar z)=e^{-1}$, i.e., a probability on the transition between the
two extremes. At $\bar z$, we have from (\ref{transcend}) that $\theta\approx 1$
(cf.\ Figure \ref{thetaXz}) and from (\ref{transcend}) and (\ref{bigknots1})
that
\begin{equation}
1\approx \bar zn\theta^2(1+\theta)(1-\theta)=
\bar zn\theta^2(1+\theta)e^{-\bar z\theta}.
\end{equation}
Consequently,
\begin{equation}
\label{zbar}
\bar z-\ln\bar z\approx\ln(2n).
\end{equation}
For very large $\bar z$, (\ref{zbar}) implies
\begin{equation}
\bar z\approx\ln(2n),
\end{equation}
but otherwise it is worthy to attempt a better solution to (\ref{zbar}). We do
so by letting $\bar z=\ln(2n)+\epsilon$ in (\ref{zbar}), which yields
\begin{equation}
\epsilon\approx\frac{\ln[\ln(2n)]}{1-\frac{1}{\ln(2n)}}
\end{equation}
and
\begin{equation}
\label{zbar1}
\bar z\approx\ln(2n)\left[1+\frac{\ln[\ln(2n)]}{\ln(2n)-1}\right].
\end{equation}

\subsection{Results of simulations}\label{new:simulation}

In this section we present additional simulation results on random digraphs
and compare them with the analytic predictions of Section \ref{new:analytic}
concerning cycle components and knots. For $n$ and $z$ fixed, each simulation
run follows the generic pattern described in Section \ref{old:simulation}. Also
as in that section, a tradeoff exists between the value of $n$ and the number of
repetitions over which the quantities of interest are averaged. We employ
different combinations, depending on the particular structure under
consideration.

\begin{figure}
\centering
\begin{tabular}{c@{\hspace{0.3in}}c}
\epsfig{file=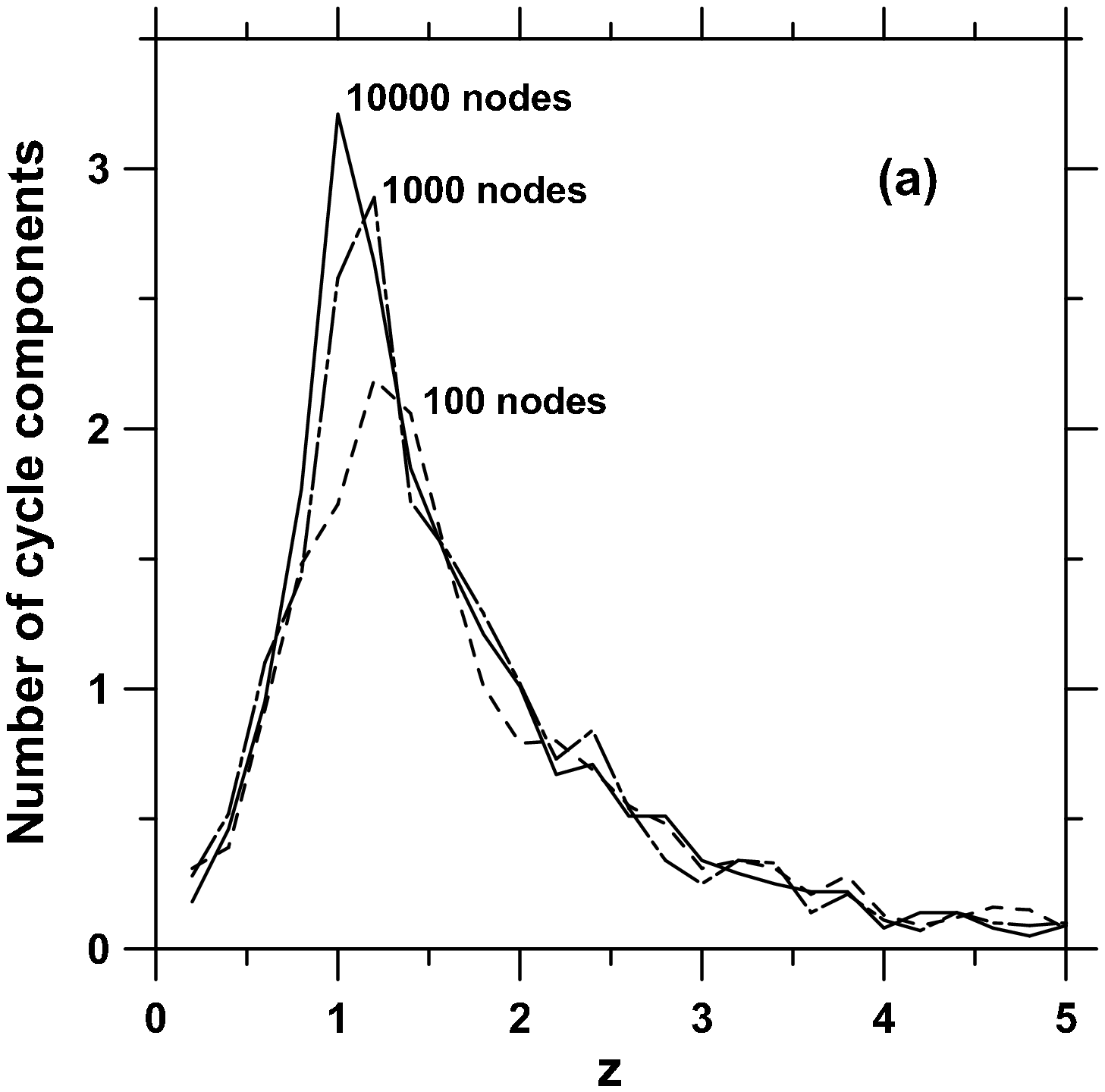, height=2.25in}&\epsfig{file=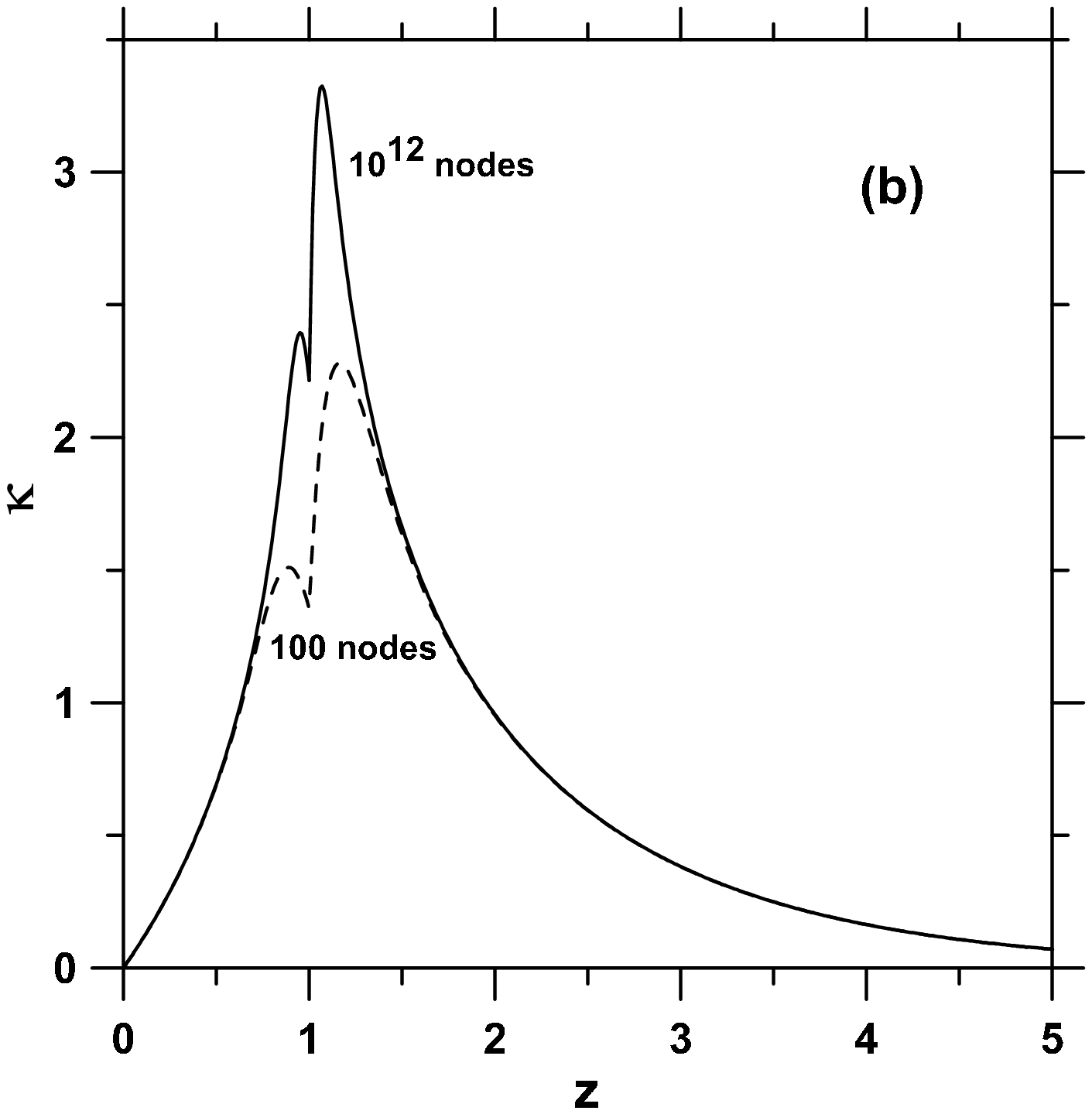, height=2.25in}\\
\end{tabular}
\begin{tabular}{c}
\epsfig{file=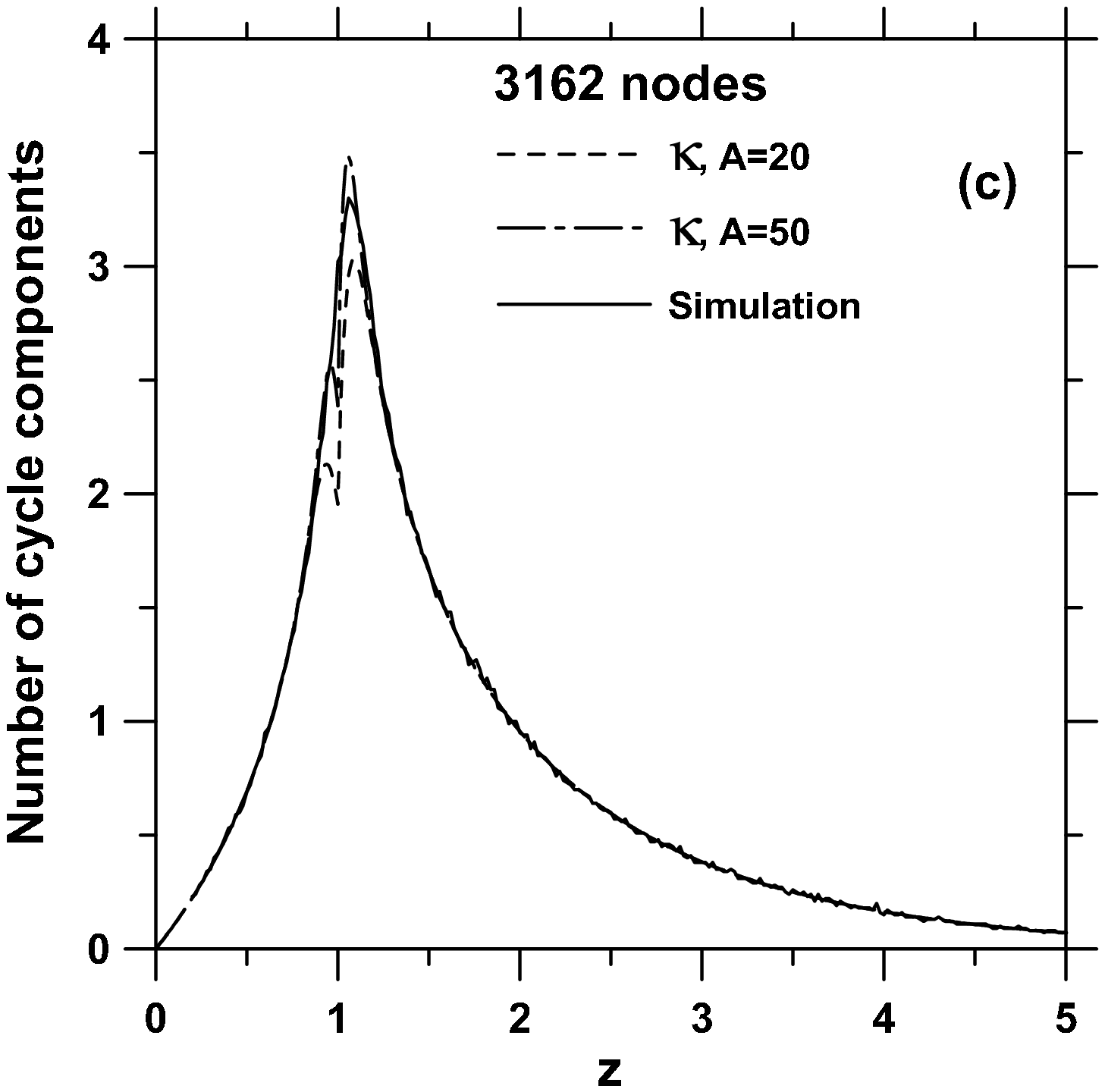, height=2.25in}\\
\end{tabular}
\caption{Average number of cycle components as a function of $z$ ($100$
simulation runs) (a); $\kappa_{z<1}(z)$ and $\kappa_{z>1}(z)$ as given by
(\ref{nczlesst1}) and (\ref{nczgreatert1}) for $A=10$ (b); average number of
cycle components as a function of $z$ ($3000$ simulation runs) and the outcome
of (\ref{nczlesst1}) and (\ref{nczgreatert1}) for two values of $A$ (c).}
\vspace{0.30in}
\label{cycleXz}
\end{figure}

We start by illustrating the existence of cycle components as a function of $z$,
for $z\le 5$. This is shown in Figure \ref{cycleXz}(a), which depicts for three
different values of $n$ the average number of cycle components found over $100$
simulation runs. All three curves peak slightly to the right of $z=1$ and decay
rapidly to either side, more markedly so for $z<1$. Our analytic prediction for
this quantity is given as a function of $A$ by $\kappa_{z<1}(z)$ in
(\ref{nczlesst1}) and $\kappa_{z>1}(z)$ in (\ref{nczgreatert1}), respectively
for $z<1$ and $z>1$. These two functions are plotted in Figure \ref{cycleXz}(b)
for $A=10$ and two values of $n$ ten orders of magnitude apart from each other.

In Figure \ref{cycleXz}(b), note first that the value of $A$ is not particularly
well suited to quantitatively reproduce what takes place for the two values of
$n$. It is, however, small enough for the computation of (\ref{nczgreatert1})
to be kept within reasonable time bounds for $n$ as large as $10^{12}$,
especially if we consider the double summation in (\ref{prob1}). The interest
in the two plots for the same value of $A$ is that they reveal that the
predicted expected number of cycle components grows very slowly with $n$, and
even so only around its peak near $z=1$. In qualitative terms, this is already
present in the simulation results shown in Figure \ref{cycleXz}(a).

For $n=3162$, Figure \ref{cycleXz}(c) shows three curves representing the
expected number of cycle components. One curve gives the average of $3000$
simulation runs, while the other two are plots of (\ref{nczlesst1}) and
(\ref{nczgreatert1}) for $A=20$ and $A=50$. It is clear from the figure that,
if $A$ were to be continuously increased from $A=20$ to $A=50$, then a value
for $A$ would certainly be found for which the two curves (the one from the
simulation and the analytic) would match nearly perfectly. The single
noteworthy exception would occur, as is apparent both in Figure \ref{cycleXz}(b)
and in Figure \ref{cycleXz}(c), in the vicinity of $z=1$. But this is really to
be expected, because all our analytic results are based on the Poisson branching
process discussed in Section \ref{old:analytic}, which fails for $z=1$ due to
the singularity that is evident in both (\ref{zlesst1}) and (\ref{zgreatert1}).

\begin{figure}
\centering
\begin{tabular}{c@{\hspace{0.3in}}c}
\epsfig{file=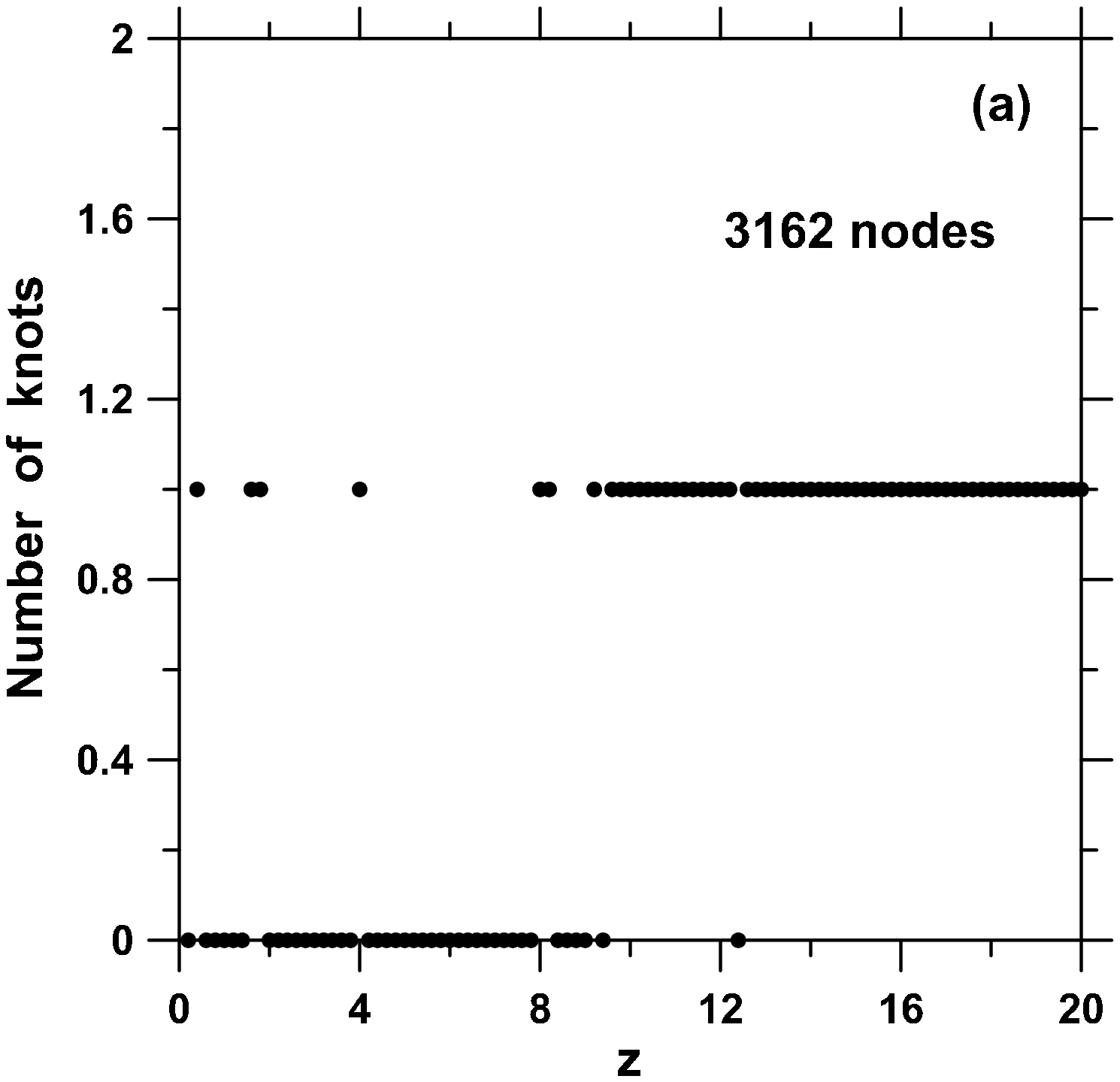, height=2.25in}&\epsfig{file=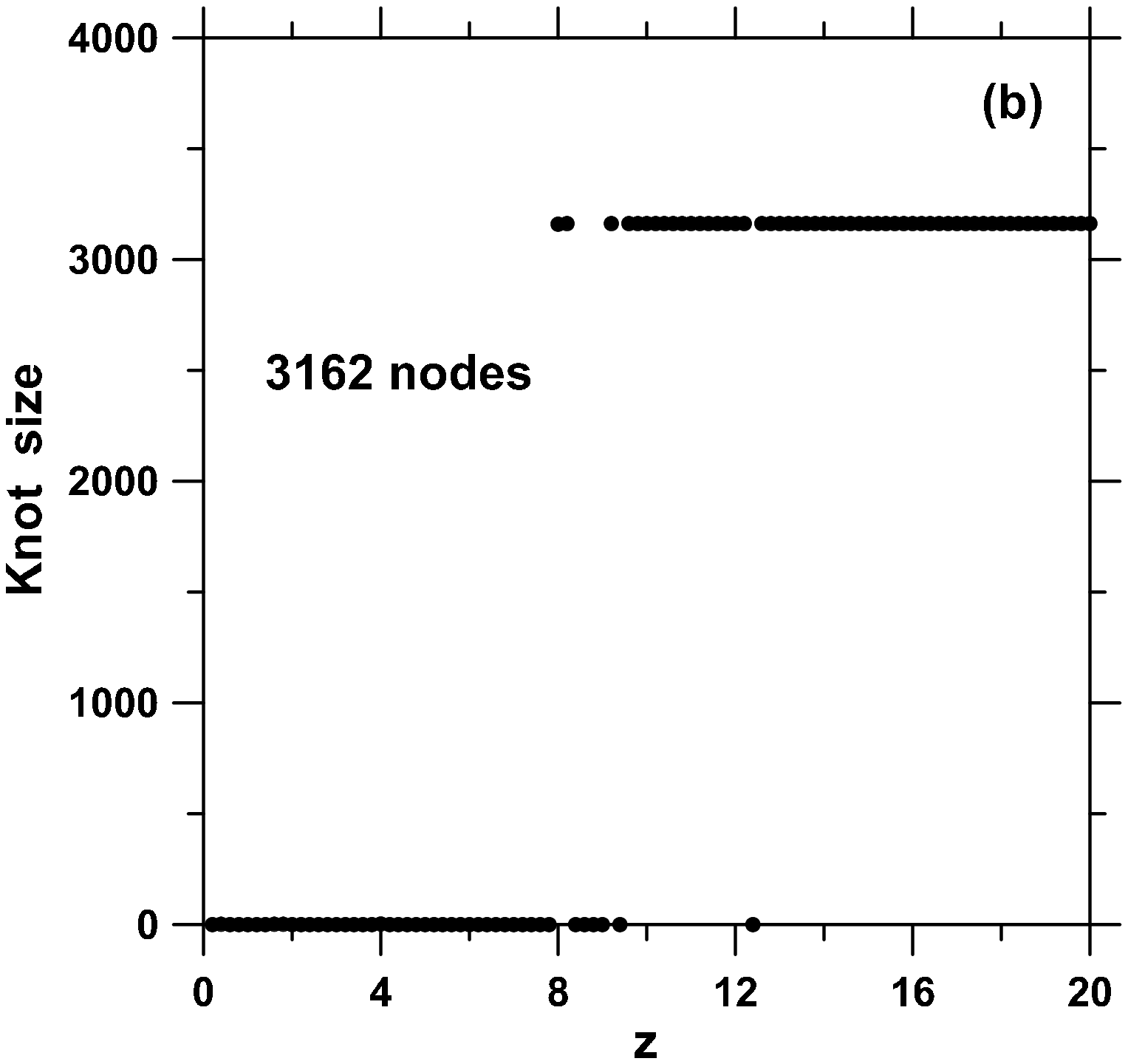, height=2.25in}\\
\epsfig{file=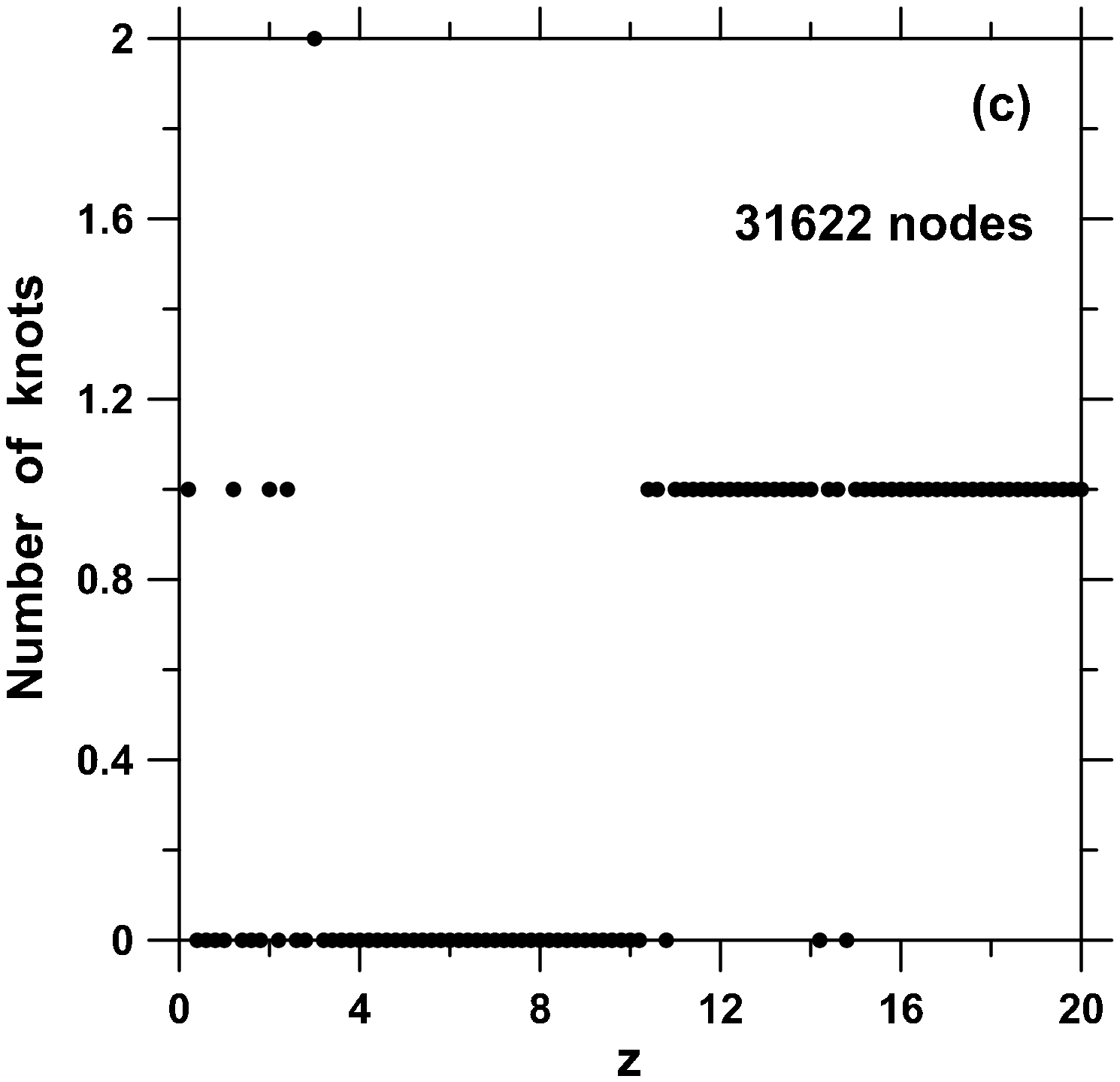, height=2.25in}&\epsfig{file=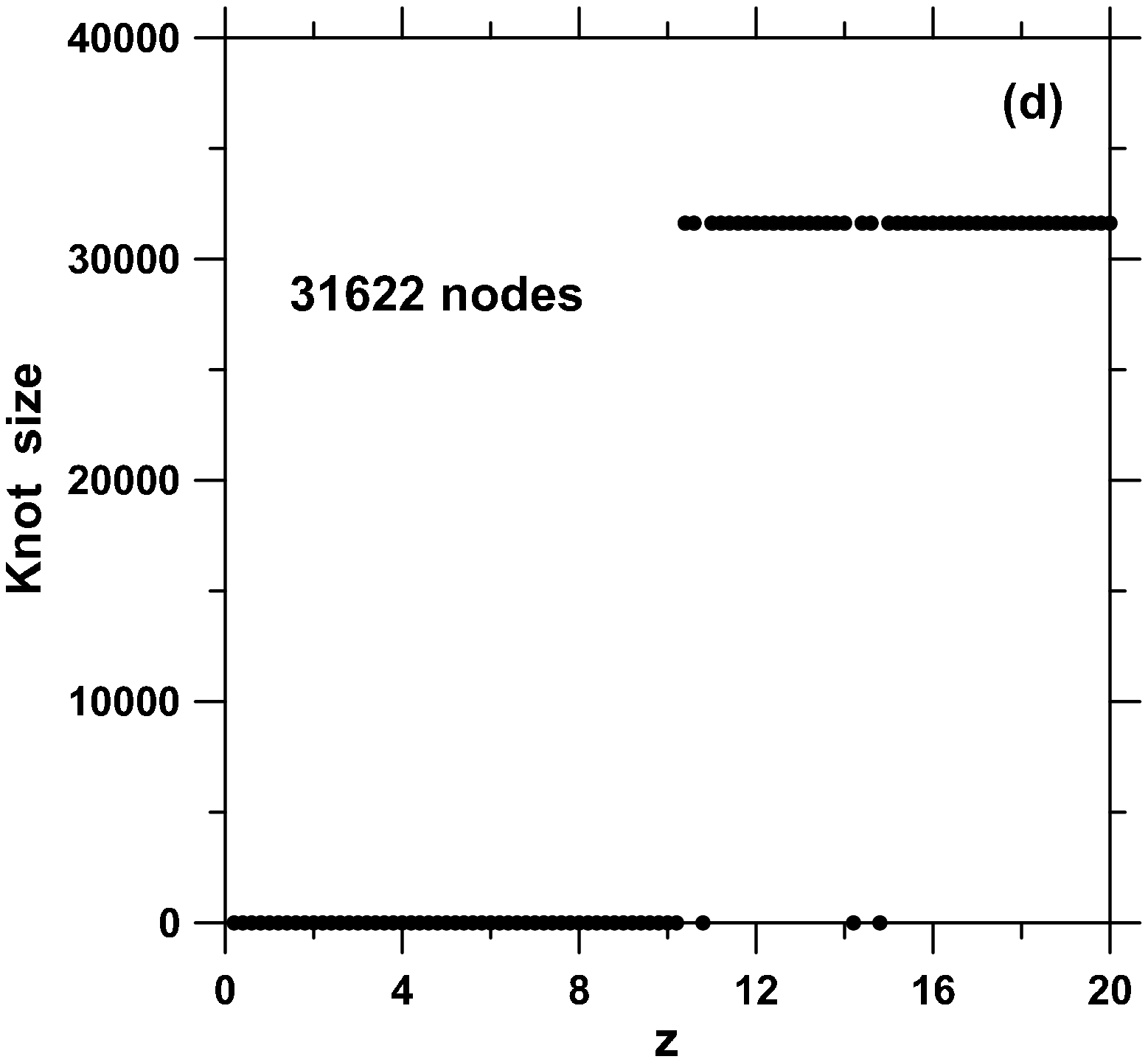, height=2.25in}\\
\end{tabular}
\caption{Number of knots as a function of $z$ (a, c), and knot size as a
function of $z$ (b, d) (one single simulation run).}
\vspace{0.30in}
\label{knotXz}
\end{figure}

The case of knots, as we know from Section \ref{new:analytic}, is considerably
simpler. In fact, for $n=3162$ and $n=31622$ we have in Figures \ref{knotXz}(a)
through \ref{knotXz}(d) a confirmation of the analytic predictions of that
section. In order to avoid the smoothing effect of averaging out over multiple
simulation runs that in this case would be undesirable, each of the four figures
depicts the result of one single simulation run for $z\le 20$. Figures
\ref{knotXz}(a) and \ref{knotXz}(c), respectively for $n=3162$ and $n=31622$,
show the number of knots as a function of $z$. As expected, either knots are
nearly inexistent or only one knot exists; the transition from the former
scenario to the latter is sharp and occurs around the value $\bar z$ of $z$
given by (\ref{zbar1}) for each $n$.

Figures \ref{knotXz}(b) and \ref{knotXz}(d) complement the picture of what
happens to knots, respectively for $n=3162$ and $n=31622$. To the left of the
aforementioned transition, the single knot that in very few cases exists has
very small size (size one, in fact, though this cannot of course be seen in the
figure). To the right of the transition, the single knot invariably encompasses
all $n$ nodes (the few points in this region for which a small value appears
correspond to inexistent knots according to Figures \ref{knotXz}(a) and
\ref{knotXz}(c), so the small value is really zero and carries no further
meaning).

\section{Conclusions}\label{concl}

We have in this paper considered random graphs and digraphs whose edges exist
with fixed probability independently for all pairs of nodes. After reviewing
some of the main known results for such graphs, notably the appearance of the
giant component (giant strong component, for digraphs), we proceeded to an
investigation of special classes of strong components in random digraphs,
specifically cycle components and knots, for which we gave analytic and
simulation results.

Our main contribution has been a detailed analytic study of the cycle
components. For these components we employed a variation of the Poisson
branching processes commonly used in the analysis of random graphs and
digraphs and derived expressions that yield the expected number of cycle
components for given $n$ and $z$. These expressions depend on the parameter
$A$ that characterizes the so-called small strong components of random digraphs,
and also on the distribution of the sum of a fixed number of Borel-distributed
random variables, which we also derived.

We believe our work also contributes in helping shift the focus of
random-graph studies toward the case of digraphs, which clearly is the case
that contemplates several of the most important random networks of current
interest. When compared to the case of random graphs, it is fair to state that
random digraphs have so far been largely neglected.

\ack

The authors acknowledge partial support from CNPq, CAPES, the PRONEX initiative
of Brazil's MCT under contracts 41.96.0857.00 and 41.96.0886.00, and a FAPERJ
BBP grant.

\bibliography{rgraph}
\bibliographystyle{elsart-num}

\end{document}